\documentclass[cmp]{svjour}
\usepackage{latexsym}
\usepackage{graphics}
\usepackage{latexsym}
\usepackage{amsmath}
\usepackage{amsfonts}
\usepackage{geometry}
\usepackage{amssymb}

\newtheorem{teo}{Theorem}[section]
\newtheorem{lem}[teo]{Lemma}
\newtheorem{corol}[teo]{Corollary}

\newtheorem{prop}[teo]{Proposition}
\newtheorem{rem}[teo]{Remark}

\newcommand\e{{\rm e}}
\newcommand\p{{\partial}}
\newcommand{\beq}{\begin{equation}}
\newcommand{\eeq}{\end{equation}}

\newcommand{\N}{{\mathbb{N}}}
\newcommand{\Z}{{\mathbb{Z}}}
\newcommand{\R}{{\mathbb{R}}}

\newcommand{\z}{{\mathcal{z}}}

\newcommand{\Res}{{\rm Res}}
\renewcommand{\Re}{{\rm Re}}

\begin{document}

\title{Spectral analysis and zeta determinant on the deformed spheres
}

\author{M. Spreafico\inst{1}\fnmsep \inst{2} \and S. Zerbini\inst{3}
}                     
\institute{ICMC-Universidade de S\~ao Paulo, S\~ao Carlos,
Brazil, \\mauros@icmc.usp.br \and \emph{Partially supported by
FAPESP: 2005/04363-4} \and Dipartimento di fisica, Universit\'a
di Trento, Gruppo Collegato di Trento, Sezione INFN di Padova, Italy.
\\zerbini@science.unitn.it}
\date{Received: date / Accepted: date}
%
\communicated{name}
\maketitle

\begin{abstract} We consider a class of singular Riemannian
manifolds, the deformed spheres $S^N_k$, defined as the classical
spheres with a one parameter family $g[k]$ of singular Riemannian
structures, that reduces for $k=1$ to the classical
metric. After giving explicit formulas for the eigenvalues and eigenfunctions of the
 metric Laplacian $\Delta_{S^N_k}$, we study the
associated zeta functions $\zeta(s,\Delta_{S^N_k})$. We introduce
a general method to deal with some classes of simple and double
abstract zeta functions, generalizing the ones appearing in
$\zeta(s,\Delta_{S^N_k})$. An application of this method allows
to obtain the main zeta invariants for these zeta functions in
all dimensions, and in particular $\zeta(0,\Delta_{S^N_k})$
and $\zeta'(0,\Delta_{S^N_k})$. We give explicit formulas for the
zeta regularized determinant in the low dimensional cases,
$N=2,3$, thus generalizing a result of Dowker \cite{Dow1}, and we compute the first coefficients in the expansion of
these determinants in powers of the deformation parameter $k$.
\end{abstract}

\section{Introduction}
\label{s0}

In the last decades there has been a (continuously increasing)
interest in the problem of obtaining explicit information on the
zeta regularized determinant of differential operators \cite{ABP}
\cite{RS} \cite{Haw} \cite{Sar} \cite{KV} \cite{Vor}. Despite the
lack of a general method, a lot of results are available in the
literature for various particular cases or by means of some kind
of approximation. Moreover, quite complete results have been
obtained for the geometric case of the metric Laplacian on a
Riemannian compact manifold for some classes of simple spaces:
spheres \cite{CQ} \cite{Cam}, projective spaces \cite{Spr1}, balls
\cite{BGKE}, orbifolded spheres \cite{Dow1}, compact (and non compact) hyperbolic manifolds
\cite{cognola92} \cite{bytsenko96} \cite{cognola97} or in
particular cases: Sturm operators on a line segment \cite{BFK}
\cite{Les}, cone on a circle \cite{Spr3}.

In particular, many
works in the recent physical literature applied this zeta function
regularization process to study the modifications induced at
quantum level by some kind of deformation of the back ground space
geometry of  physical models \cite{hu} \cite{Dow} \cite{CD} \cite{SV}.
 In this context, a full class of deformed spaces, called
deformed spheres, has been introduced in \cite{SV}, where the
perturbation of the heat kernel expansion has been studied. This  
is a particularly interesting class of spaces in Einstein theory
of gravitation  and in  cosmology, since the appearance of a non
trivial deformation produces a  symmetry braking of the space. In
fact, the deformed sphere may be considered as the Euclidean
version of the a deformed de Sitter space, which is particular
relevant in modern cosmology, since it represents the
inflationary  as well as the recent accelerated phase. It is well
known that the quantum effective action is related to the
regularized functional determinant of Laplace type operators
(see, for example \cite{EORBZ} and references therein). As a
consequence, an  expansion of functional determinant with respect
to deformation parameter around its spherical symmetric value
describes the effects of such geometric symmetry breaking.

It is therefore a natural question to see if the explicit
calculation of the zeta determinant for the Laplace operator on
this class of spaces is possible. 
In this work we give a positive
answer to this question, establishing a general method that
permits to compute the zeta regularized determinant on a deformed
sphere of any dimension. Actually, for a particular discrete set of values of the deformation parameter $k$, the $N$-dimensional deformed sphere turns out to be isometric to the so called orbifolded sphere, the quotient space $S^N/\Gamma$, of the standard $N$-sphere by a finite subgroup of the rotation group $O_N(\R)$. Determinants on these spaces have been studied by J.S. Dowker in a series of works \cite{Dow1} \cite{Dow2} \cite{Dow3}, where results are also  obtained for different couplings. 
Under this point of view, the present work is a generalization of the results of Dowker to the contiunous range of variation of the deformation parameter $k$, and in fact the results are consistent (see Section \ref{s3}).

The main motivation of the present work, beside the particular result, is that the method
introduced  has the advantage of being completely general and not
related to this specific problem. In particular, we show how it
can be applied to obtain the main zeta invariants of some classes
of abstract simple and double zeta functions (Sections \ref{s3.2}
and \ref{s3.3}).
In order to give the explicit form for the zeta function on the
deformed spheres, we produce an explicit description of the spectrum
and of the eigenfunctions of the associated Laplace operator in
any dimension (Proposition \ref{s2.p2}). In particular, the 2
dimensional case turns out to be very interesting, from the point of view of geometry: in fact the 2
dimensional deformed sphere is a space with singularities of
conical type. This class of singular spaces was introduced and
studied by Cheeger \cite{Che} and although since them became a
subject of deep interest and investigation, there are in fact
relatively few occasions where explicit results can be obtained.

\section{The geometry of the deformed spheres}
\label{s1}

In this section we provide the definition of the $N$ dimensional
deformed sphere $S_k^N$, where $k$ is the deformation parameter,
and we study its geometry.  This produces a particular
interesting relation with elliptic function and conical
singularity, at least in the 2 dimensional case. 
The deformed
$N$-sphere is defined as the standard $N$-sphere with a singular
Riemannian structure. When $N=2$, we have an isometry with the
surface immersed in $\R^3$ that can be obtained by rotating around
an axis a curve described by an elliptic integral function. The
surface obtained, presents two singular points of conical type, as
considered by Br\"{u}ning and Seeley in \cite{BS} generalizing the
definition of metric cone of Cheeger \cite{Che}. Thus, the 2
dimensional deformed sphere is a space with singularities of
conical type, and due to the great interest in this kind of
singular space, both from the point of view of differential geometry and
zeta function analysis (see for example \cite{Dow0} \cite{F}
\cite{C} \cite{BDK} \cite{Z} \cite{CZ}), its study is of
particular interest (compare also with \cite{Spr3}).

Consider the immersion of the $N+1$ dimensional sphere $S^{N+1}$
in $\R^{N+2}$
\[
\left\{\begin{array}{l} x_0= \sin\theta_0 \sin \theta_1\dots
\sin\theta_N\\
x_1= \sin\theta_0 \sin \theta_1\dots
\cos\theta_N\\
\dots\\
x_{N+1}=\cos\theta_0,
\end{array}
\right.
\]
and the induced metric (in local coordinates)
$g_{S^{N+1}}=(d\theta_0)^2+\sin^2\theta_0 g_{S^N}$. We deform this
metric as follows. Let $k$ be a real parameter with $0< k\leq 1$,
and consider the family
\[
g_{S^{N+1}}[k]=(d\theta_0)^2+ \sin^2\theta_0 g_{S^N}[k],
\]
\[
g_{S^1}[k]=k^2(d\theta_0)^2.
\]

This is a one parameter family of singular Riemannian metric on
$S^{N+1}$. We call the singular Riemannian manifolds
$(S^{N+1},g_{S^{N+1}}[k])$ the deformed spheres of dimension $N+1$
and we use the notation $S_k^{N+1}$. By direct inspection, we see
that the locus of the singular points of the metric in dimension
$N+1$ is a sub manifold isomorphic to two disjoint copies of
$S^{N-1}$. 
In particular, in the 2 dimensional case we have
\[
g_{S^{2}}[k]=(d\theta_0)^2+k^2\sin^2\theta_0 (d\theta_1)^2,
\]
that shows that the deformed 2-sphere $S_k^2$ is a space with
singularities of conical type as defined in \cite{BS}. 
Proceeding as in \cite{BS} Section 7, we will show in
the next subsection that the singularity is generated by rotation
of a curve in the plane.

Observe that, in a different language, $S^2_k$ is a periodic {\it lune}, 
that is to say it can be pictured by taking a segment of the standard $2$-sphere (a lune) and identifying the sides. 
This situation generalizes to higher dimensions \cite{AD}, and when the angle of the lune is $\frac{\pi}{n}$, $n\in \Z$, we obtain a spherical orbifold $S^N/\Gamma$, as pointed out in the introduction.

Note also that, by direct verification on the local
description of the metric $g_{S^{N}}[k]$, the non compact
Riemannian manifold obtained by subtracting the singular subspace
of the metric from $S^{N}_k$, is a space of constant curvature and
locally symmetric. It is not symmetric, as it is clear from the
geometry of the low dimensional cases, or observing that it is not
simply connected (see Corollary 8.3.13 of \cite{wolf}). On the
other side, the classical sphere $S^{N}_1$ is a symmetric space;
for example, the 2 dimensional one having the maximum number
$\frac{N(N+1)}{2}$ of global isometries, namely the 3 spatial
rotations. Therefore, the variation of the parameter $k$ away from
the trivial value produces a breaking of the global symmetric type
of the space. In particular for example on the 2 sphere it breaks
two continuous rotations in one discrete symmetry, namely the
reflection through the horizontal plane.

We conclude this subsection with the explicit expression for the
Laplace operator. With $a=\frac{1}{k}$, the (negative) of the
induced Laplace operator on the deformed sphere $S^{N+1}_k$ is
\[
\Delta_{S^{N+1}_{k}}=-d_{\theta_0}^2-N\frac{\cos\theta_0}{\sin\theta_0}d_{\theta_0}
+\frac{1}{\sin^2\theta_0}\Delta_{ S_k^N}.
\]

\subsection{Elliptic integrals and the deformed 2-sphere}

The geometry of the 2 dimensional case is particularly interesting
and this subsection is dedicated to its study. The ellipse
$x^2+\frac{y^2}{b^2}=1$, can be given parametrically in the first
quadrant by the formula
\[
\left\{\begin{array}{l}x=t,\\y=b\sqrt{1-t^2},\end{array}\right.
\]
where $0\leq t\leq 1$. If we assume $b\leq 1$, the arc length is
\[
l(t)=\int_0^t \sqrt\frac{1-k^2 s^2}{1-s^2}ds.
\]
where $k=\sqrt{1-b^2}$. With the new variables $t=\sin\theta$,
$s=\sin\psi$, we obtain
\[
\left\{\begin{array}{l}x=\sin\theta,\\y=b\cos\theta,\end{array}\right.
\]
with $0\leq\theta\leq\frac{\pi}{2}$, and  the arc length is
\[
E(\theta,k)=l(\sin\theta)=\int_0^{\theta} \sqrt{1-k^2
\sin^2\psi}d\psi,
\]
that is the elliptic integral of the second kind in Legendre
normal form \cite{GZ} 8.110.2 (see \cite{PS} or \cite{WW} for
elliptic functions and integrals). Note that we cannot find a
parameterization of the curve by the arc length reversing the above
equation using Jacobi elliptic functions. Consider now the curve
$f(\sin\theta)=E(\theta,k)$. This is a smooth curve in the
interval $0\leq t\leq 1$, with $f(0)=0$ and
$f(1)=E(\frac{\pi}{2},k)$. We can rotate this function around the
horizontal axis getting a surface with a geometric singularity at
the origin. For further use, it is more convenient to place the
surface in the upper half space. Thus, we consider the function
\[
f(t)=E(\arccos\frac{t}{k},k)=\int_0^{\sqrt{1-\frac{t^2}{k^2}}}\sqrt\frac{1-k^2s^2}{1-s^2}
ds,
\]
with $0\leq t\leq k$, and the curve: $x=t$, $y=f(t)$.
We reparametrize this curve by its arc length
\[
\theta=l(t)=\int_0^t
\sqrt{1+(y'(s))^2}ds=\int_0^t\frac{1}{\sqrt{k^2-s^2}}
ds=\arcsin\frac{t}{k},
\]
obtaining
\[
\left\{\begin{array}{l}x=k \sin\theta,\\y=f(k\sin
\theta)=E(\frac{\pi}{2}-\theta,k)=\int_0^{\frac{\pi}{2}-\theta}\sqrt{1-k^2\sin^2\psi}d\psi,\end{array}\right.
\]
with $0\leq \theta\leq \frac{\pi}{2}$ (as before $\theta$ is the
angle from the vertical axis).

Let now consider the surface $Y^+_k$ obtained by rotating the
above curve along the vertical axis. We have the parameterization
\[
Y_k^+:\left\{\begin{array}{l}x=k\sin\theta\cos\phi,\\y=k\sin\theta
\sin\phi,\\z=E(\frac{\pi}{2}-\theta,k),\end{array}\right.
\]
where $0\leq\phi\leq 2\pi$. This is clearly a smooth surface
except at the possible singular point $(0,0,E(\pi/2,k)))$, with
the circle $C_k$: $x^2+y^2=k^2$, $z=0$ of radius $k$ as boundary.
Moreover, since the coordinate line tangent vectors on the
boundary are $v_\phi=-\sin\phi e_x+\cos\phi e_y$ and
$v_\theta=e_z$, the tangent space is vertical and hence we can
glue smoothly $Y_k^+$ with the surface $Y_k^-$ obtained by
reflecting through the horizontal plane. We call the surface
obtained $Y^2_k=Y_k^+ \cup Y_k^-$, and the parameter $k$
deformation parameter. The surface $X_k$ obtained from $Y^2_k$ by
removing the poles $(0,0,\pm k)$ is clearly a smooth (non compact)
surface. The Riemannian metric induced on $X_k$ from the immersion
in $\R^3$ is
\[
g_{Y_k^2}(\theta,\phi)=(d\theta)^2+k^2 \sin^2\theta (d\phi)^2.
\]

It is clear that the local map $f:(\theta,\phi)\mapsto (\theta,\phi)$,
extends to a diffeomorphism $f:Y^2_k\to S^2$, and since $f^*
g_{S^2}[k]=g_{Y_k^2}$, it follows that $f$ is an isometry between
$S^2_k=(S^2,g_{S^2}[k])$ and $(Y_k^2,g_{Y_k^2})$.

\section{Spectral analysis}
\label{s2}

In this section we give the
eigenvalues and eigenfunctions of the Laplace operator on the
deformed sphere. As observed in Section \ref{s1}, the two
dimensional case  is of particular interest, since it represents an
instance of a space with singularities of conical type that can be
solved explicitly. Therefore we spend a few words to describe the concrete operator appearing in that case, using the language of spectral analysis for spaces with conical singularities \cite{Che} \cite{BS}.
With $a=\frac{1}{k}$, the (negative) of the induced Laplace
operator on the deformed sphere $S^2_k$ is
\[
\Delta_{S^2_{1/a}}=-\p_{\theta}^2-\frac{\cos\theta}{\sin\theta}\p_{\theta}-\frac{a^2}{\sin^2\theta}\p_\phi^2,
\]
on $L^2(S^2_{1/a})$. With the Liouville transform $u=Ev$, with
$E(\theta)=\frac{1}{\sqrt{\sin\theta}}$, we obtain the operator
\[
L_a=-\p_{\theta}^2-\frac{a^2}{\sin^2\theta}\p_\phi^2-\frac{1}{4}\left(1+\frac{1}{\sin^2\theta}\right).
\]

This is a regular singular operator as defined in \cite{BS},
\[
L_a=-d_\theta^2+\frac{1}{\theta^2}A(\theta),
\]
where
\[
A(\theta)=\frac{\theta^2}{\sin^2\theta}\left(-a^2
\p_\phi^2-\frac{1}{4}\right)-\frac{1}{4}\theta^2,
\]
is a family of operators on the section of the cone, that is the
circle of radius 1. It is clear that the operator $-\p_\phi^2$ has
the complete system $\left\{\mu_m=m^2,\e^{i m\phi}\right\}$, with
$m\in \Z$, where all the eigenvalues are double up to the null one
that is simple with the unique eigenfunction given by the constant
map. Since the problem decomposes spectrally on this system, we
reduce to study the family of singular Sturm operators
\[
T_{am}=-d_{\theta}^2+\frac{a^2m^2-\frac{1}{4}}{\sin^2\theta}-\frac{1}{4}.
\]

In order to define
an appropriate self adjoint extension, we introduce the following
boundary conditions at the singular points:
\[
BC_0:~~~~~~ \lim_{\theta\to 0}\theta^{am}\left[\left(a
m+\frac{1}{2}\right)\frac{1}{\theta}f(\theta)-\sqrt{\theta}f'(\theta)\right]=0,
\]
and
\[
BC_\pi:~~~~~~ \lim_{\theta\to \pi}(\theta-\pi)^{am}\left[\left(a
m+\frac{1}{2}\right)\frac{1}{\theta-\pi}f(\theta-\pi)-\sqrt{\theta-\pi}f'(\theta-\pi)\right]=0.
\]

These are the natural generalizations of the classical Dirichlet
boundary conditions (compare with \cite{Weid} 8.4) and were first
considered in \cite{BS}. In particular, it was proved in
\cite{BS}, Section 7, that the self adjoint extension defined by
these conditions is the Friedrich extension.

The eigenvalues equation associated with the operators
$T_{am}$, can be more easily studied going back to the original
Hilbert space. This equation was in fact already studied by Gromes \cite{Gro}, who found a complete solution. Generalizing the standard approach used for the standard sphere (see for example \cite{Hob}), we can prove that in fact this solution provide a complete set of eigenvalues and eigenfunctions for the metric Laplacian, as stated in the following lemma.

\begin{lem} The operator $\Delta_{S^2_{1/a}}$, has the complete system:
\[
\lambda_{n,m}=(a|m|+n)(a|m|+n+1), ~n\in \N, m\in \Z,
\]
where all the eigenvalues with $m\not=0$ are double with
eigenfunctions (where the $P_\nu^\mu$ are the associated Legendre functions)
\[
\e^{iam\phi}P^{-am}_{am+n}, ~~\e^{-iam\phi}P^{-am}_{am+n},
\]
while the eigenvalues $n(n+1)$ are simple with eigenfunctions the
functions $P_n$. \label{s2.p1}
\end{lem}

Next, we pass to the higher dimensions. The (negative) of the induced Laplace operator on the deformed
sphere $ S^{N+1}_k$ is
\[
\Delta_{
S^{N+1}_{k}}=-d_{\theta_0}^2-N\frac{\cos\theta_0}{\sin\theta_0}d_{\theta_0}
+\frac{1}{\sin^2\theta_0}\Delta_{ S_k^N}.
\]
on $L^2( S^{N+1}_k)$. Projecting on on the spectrum of $\Delta_{
S_k^N}$, we obtain the differential equation
\[
\left[-d_{\theta_0}^2-N\frac{\cos\theta_0}{\sin\theta_0}d_{\theta_0}
+\frac{\lambda_{
S^N_k}}{\sin^2\theta_0}\right]u=\lambda_{S_k^{N+1}} u.
\]

Following \cite{SV}, we make the substitutions
\[
u(\theta_0)=\sin^b\theta_0 v(\theta_0),
\]
\[
z=\frac{1}{2}(\cos\theta_0+1),
\]
where $b=\frac{1}{2}\left(1-N+\sqrt{(N-1)^2+4\lambda_{
S^N_k}}\right)$. This gives the hypergeometric equation \cite{GZ}
9.151
\[
z(1-z)v''+[\gamma-(\alpha+\beta+1)z]v'-\alpha\beta v=0,
\]
with
\[
\alpha=\frac{1}{2}\left(2b+N\mp\sqrt{N^2+4\lambda_{
S^{N+1}_k}}\right),
\]
\[
\beta=\frac{1}{2}\left(2b+N\pm\sqrt{N^2+4\lambda_{
S^{N+1}_k}}\right),
\]
\[
\gamma=\frac{1}{2}(2b+N+1).
\]

Boundary conditions give the equation
\[
2n+2b+N=\pm\sqrt{N^2+4\lambda_{ S^{N+1}_k}},
\]
where $n\in \N$, that, in turns, gives the recurrence relation
\[
\lambda_{S^{N+1}_k}=n^2+\left(1+\sqrt{(1-N)^2+4\lambda_{
S^{N}_k}}\right)n+\frac{1}{2}\left(1-N+\sqrt{(1-N)^2+4\lambda_{
S^{N}_k}}+2\lambda_{ S^{N}_k}\right).
\]

We can prove that this recurrence relation is satisfied by the
numbers
\[
\lambda_{S^{N}_k}=\left(am+n_1+\dots+
n_{N-1}+\frac{N-1}{2}\right)^2-\frac{(N-1)^2}{4},
\]
where $n_i\in \N$, must be  a positive integer. We have obtained
\[
b=am+n_1+\dots+n_{N-1},
\]
\[
\alpha=-n_N,
\]
\[
\beta=2(am+n_1+\dots+n_{N-1})+n_N+N,
\]
\[
\gamma=am+n_1+\dots+n_{N-1}+\frac{N+1}{2},
\]
and the family of solutions for the eigenvalues equation
(up to a constant)
\[
u_{n_N}(\cos\theta_0)=\sin^\frac{1-N}{2}\theta_0
P^{-am-n_1-\dots-n_{N-1}-\frac{N-1}{2}}_{am+n_1+\dots+n_{N-1}+\frac{N-1}{2}+n_N}(\cos\theta_0).
\]

Using standard argument, we can then prove the following result.

\begin{lem} The operator $\Delta_{ S^{N+1}_{1/a}}$, has the complete system:
\[
\lambda_{m,n_1,\dots,
n_N}=(a|m|+n_1+\dots+n_N)(a|m|+n_1+\dots+n_N+N), ~n_i\in \N, m\in
\Z,
\]
where all the eigenvalues with $m\not=0$ are double with
eigenfunctions (up to normalization)
\[
\e^{iam\theta_N}\prod_{j=0}^{N-1}\sin^\frac{1-N+j}{2}(\theta_j)
P^{-am-n_1-\dots-n_{N-1-j}-\frac{N-1-j}{2}}_{am+n_1+\dots+n_{N-j}+\frac{N-1-j}{2}}(\cos\theta_j),
\]
\[
\e^{-iam\theta_N}\prod_{j=0}^{N-1}\sin^\frac{1-N+j}{2}(\theta_j)
P^{-am-n_1-\dots-n_{N-1-j}-\frac{N-1-j}{2}}_{am+n_1+\dots+n_{N-j}+\frac{N-1-j}{2}}(\cos\theta_j),
\]
while the eigenvalues with $m=0$ are simple with eigenfunctions
\[
\prod_{j=0}^{N-1}\sin^\frac{1-N+j}{2}(\theta_j)
P^{-n_1-\dots-n_{N-1-j}-\frac{N-1-j}{2}}_{n_1+\dots+n_{N-j}+\frac{N-1-j}{2}}(\cos\theta_j).
\]
\label{s2.p2}
\end{lem}

\section{Zeta regularized determinants}
\label{s3}

In this section we study the zeta function associated to the
Laplace operator on the deformed sphere $ S_k^{N+1}$. For, we
introduce two quite general classes of zeta functions and we
compute the main zeta invariants of them. This allows us to define
a general technique to obtain the zeta regularized determinant of
the Laplace operator on $ S^{N+1}_k$  as a function of the
deformation parameter. We apply this technique to the lower cases,
$N=1$ and $2$, giving explicit formulas. Our last result is the
computation of the coefficients in the expansions of the zeta
determinants in powers of the deformation parameter.

By Proposition \ref{s2.p2}, the zeta function on $ S^{N+1}_k$ is
the function defined by the series
\[
\zeta(s,\Delta_{ S^{N+1}_{1/a}})=\sum_{n\in \N_0^N}
[n(n+N)]^{-s}+2\sum_{m=1}^\infty \sum_{n\in
\N^N}[(am+n)(am+n+N)]^{-s},
\]
when $\Re(s)>N+1$, and by analytic continuation elsewhere. Here
$n$ is a positive integer vector $n=(n_1,\dots,n_N)$, and the
notation $\N_0^{N}$ means $\N\times\dots \times \N-\{0,\dots,0\}$.

Multidimensional Gamma and zeta functions, namely zeta functions
where the general term is of the form $(n^Tan+b^T n+c)^{-s}$,
where $a$ is a real symmetric matrix of rank $k\geq 1$, $b$ a
vector in $\R^k$, $c$ a real number and $n$ an integer vector in
$\Z^k$, were originally introduced by Barnes \cite{Bar1}
\cite{Bar2} and Epstein \cite{Eps1} \cite{Eps2} as natural
generalizations of the Euler Gamma function. Whenever the sum is
on the integers (i.e. $n\in \Z^k$), there is a large symmetry
that allows to express the zeta function by a theta series.
Multidimensional theta series have been deeply studied in the
literature, and by a generalization of the Poisson summation
formula (see for example \cite{Cha} XI.2, 3) it is possible to
compute the main zeta invariants for multiple series of this type
(see \cite{Wei} \cite{OS} \cite{EORBZ} \cite{E} and \cite{CN2}
and references thereby). The main problem in the present case is
that the zeta functions are associated to series of Dirichlet
type, namely the sums are over $\N_0^k$. We lose then many
symmetries and in particular a formula of Poisson type.
Consequently, it is more difficult to find general results, and
different techniques have been introduced to deal with the
specific cases (see for example \cite{CMB1} \cite{CMB2} \cite{CQ}
\cite{Eie2} \cite{Spr1} \cite{Spr2} for simple series or series
that can be reduced to simple series or \cite{CN2} \cite{Mat} for
multiple linear series). Note in particular that the case of a
double ($k=2$) homogeneous quadratic series of Dirichlet type is
much harder. The zeta functions  of this type (with integer
coefficients) appear when dealing with the zeta functions of a
narrow ideal class for a real quadratic field as shown by Zagier
in \cite{Zag1} and \cite{Zag2}, where he also computes the values
at non positive integers (see also \cite{Shi} \cite{Eie1}
\cite{CN1} \cite{CN2}, and in particular \cite{Spr5} for the
derivative). Beside, we can overcome this difficulty in the case under study
first by reducing the multi dimensional zeta functions to a sum of 2
dimensional linear and quadratic zeta functions, and then studying
the quadratic one by means of a general method introduced in
\cite{Spr6} in order to deal with non homogeneous zeta functions.
Note that, for particular values of the deformation parameter, the zeta function can be reduced to a sum of zeta functions of Barnes type \cite{Bar1}, and this allows a direct computation of the main zeta invariants \cite{Dow1} \cite{Dow2}. This approach does not work for generic values of the deformation parameter, and therefore the more sophisticated technique introduced here is necessary.

We present in the next subsection some generalizations of some
results of \cite{Spr6} necessary in order to treat the present
case, and we give in the following subsections some applications
to the case of some general classes of abstract simple and double
zeta functions. As explained here after, by means of these two
classes of zeta functions, we can in principle calculate the zeta
invariants for the deformed sphere in any dimensions. Eventually in
the last subsections we apply the method to obtain the main zeta
invariants for the zeta functions on the 2 and 3 dimensional
deformed spheres.

By the following lemma (see \cite{Var} or \cite{Spr6}), we can
reduce $\zeta(s,\Delta_{S_k^{N+1}})$ to a sum of simple and double
zeta functions.

\begin{lem}\label{s3.l1} Let $f(z)$ be a regular function of $z$.
Then
\[
\sum_{n\in \N^{N+1}} f(n)=\sum_{n=0}^\infty \binom{n+N}{N} f(n),
\sum_{n\in \N^{N+1}_0} f(n)=\sum_{n=1}^\infty \binom{n+N}{N} f(n).
\]
\end{lem}

\begin{prop} The zeta function associated with the Laplace operator on the
$N+1$ dimensional deformed sphere is ($N\geq 1$)
\[
\zeta(s,\Delta_{ S^{N+1}_{1/a}})=\sum_{n=1}^\infty
\binom{n+N-1}{N-1}[n(n+N)]^{-s}
+2\sum_{m=1}^\infty \sum_{n=0}^\infty
\binom{n+N-1}{N-1}[(am+n)(am+n+N)]^{-s}.
\]
\label{s3.p0}
\end{prop}

Since $\binom{n+N-1}{N-1}=P_N(n)$ is a polynomial of order $N$ in
$n$, and since given any polynomial $P_N(n)$ we have a polynomial
$Q_N(n+x)$ for any given $x$, such that $P_N(n)=Q_N(n+x)$ (and we
can find explicitly the coefficients of $Q$ as functions on those
of $P$ and $x$), it is sufficient to consider the two classes of
zeta functions
\[
z(s;\alpha,2,x,p)=\sum_{n=1}^\infty (n+x)^\alpha [(n+x)^2+p]^{-s},
\]
and
\[
Z(s;\alpha,a,x,p)=\sum_{m=1}^\infty\sum_{n=1}^\infty n^\alpha
[(n+am+x)^2+p]^{-s}.
\]

This will be done in Subsections \ref{s3.2} and \ref{s3.3}, but
first, the next subsection is dedicated to recall and generalize
some results on sequences of spectral type and associated zeta
functions introduced in \cite{Spr6}, necessary in the following.

\subsection{Sequences of spectral type and zeta invariants}
\label{s3.1}

In this subsection we will use some concepts and results developed
in \cite{Spr6}, that briefly we recall here. We refer to that work
for further details and complete proofs.

Let $T=\{\lambda_n\}_{n=1}^\infty$ be a sequence of positive
numbers with unique accumulation point at infinite, finite
exponent $s_0$ and genus $q$. We associate to $T$, the heat
function
\[
f(t,T)=1+\sum_{n=1}^\infty \e^{-\lambda_n t},
\]
the logarithmic Fredholm determinant
\[
\log F(z,T)=\log \prod_{n=1}^\infty
\left(1+\frac{z}{\lambda_n}\right)\e^{\sum_{j=1}^q
\frac{(-1)^j}{j}\frac{z^j}{\lambda_n^j}},
\]
and the zeta function
\[
\zeta(s,T)=\sum_{n=1}^\infty \lambda_n^{-s}.
\]

The sequence $T$ is called of spectral type if there exists an
asymptotic expansion of the associated heat function for small $t$
in powers of $t$ and powers of $t$ times positive integer powers
of $\log t$. In particular it is said to be a simply regular
sequence of spectral type if the associated zeta function as at
most simple poles (see \cite{Spr6} pg. 4 and 9). Formulas to deal
with the zeta invariants for sequences of spectral type are given
in \cite{Spr6}. In particular, there are considered  non
homogeneous sequences as well. We generalize the concept of non
homogenous sequence here, by considering, for any given sequence
of spectral type $T_0=\{\lambda_n\}_{n=1}^\infty$, the shifted
sequence $T_d=\{\lambda_n+d\}_{n=1}^\infty$, where $d$ is a 
parameter, subject to the unique condition that $\Re(\lambda_n+d)$ is
always positive. We can prove the following results for a shifted
sequence (see \cite{Spr6} Proposition 2.9 and Corollary 2.10 for details). 

\begin{lem} \label{s3.l2} Let $T_0=\{\lambda_n\}_{n=1}^\infty$ be a
sequence of finite exponent $s_0$ and genus $q$, then the
associated shifted sequence $T_d=\{\lambda_n+d\}_{n=1}^\infty$,
with $d$ such that $\Re(\lambda_n+d)>0$ for all $n$, is a sequence
of finite exponent $s_0$ and genus $q$. Moreover, $T_0$ is of
spectral type if and only if $T_d$ is of spectral type. If $T_0$
is simply regular, so is $T_d$.
\end{lem}

\begin{prop} Let $T_0=\{\lambda_n\}_{n=1}^\infty$ be a simply
regular sequence of spectral type with finite exponent $s_0$ and
genus $q$, and $T_d=\{\lambda_n+d\}_{n=1}^\infty$, with $d$
such that $\Re(\lambda_n+d)>0$ for all $n$, an  associated shifted
sequence. Then,
\[
\zeta(0,T_d)=\zeta(0,T_0) +\sum_{j=1}^q\frac{(-1)^j}{j}{\rm
Res}_1(\zeta(s,T_0),s=j)d^j,
\]
\[
\zeta'(0,T_d) =\zeta'(0,T_0) -\log F(d,T_0)+
\]
\[
+\sum_{j=1}^q\frac{(-1)^j}{j} \left[{\rm
Res}_0(\zeta(s,T_0),s=j)+(\gamma+\psi(j)){\rm
Res}_1(\zeta(s,T_0),s=j)\right]d^j.
\]
\label{s3.p2}
\end{prop}

\begin{prop} \label{product}  Let $T_0=\{\lambda_n\}_{n=1}^\infty$ be a simply
regular sequence of spectral type with finite exponent $s_0$ and
genus $q$. Let $L_0=\{\lambda_n^2\}_{n=1}^\infty$, and  $d$
such that ${\rm Re}(\lambda_n+d)>0$ for all $n$. Then,
\[
\zeta(0,L_{d^2})
=\frac{1}{2}\left[\zeta(0,T_{i d})+\zeta(0,T_{-i d}\right],
\]
\[
\zeta'(0,L_{d^2})
=\zeta'(0,T_{i d})+\zeta'(0,T_{-i d})
-\sum_{j=1}^{\left[\frac{p}{2}\right]}\frac{(-1)^j}{j}
\sum_{k=1}^j\frac{1}{2k-1}{\rm Res}_1(\zeta(s,T_0),s=2j) d^{2j}.
\]
\end{prop}

\begin{rem} Note that the numbers $\lambda_n$ in the sequence
need not to be different, i.e. the cases with multiplicity are
covered by Propositions \ref{s3.p2} and \ref{product}. In particular, assume the
sequence is $T_0=\{\lambda_n\}_{n=1}^\infty$, each $\lambda_n$
having multiplicity $\rho_n$ (we cover the case of a general
abstract multiplicity, given by any positive real number). Then,
the unique difficulty can be in defining the exponent of
convergence of the sequence. But actually for our purpose it is
sufficient to know the genus, and this can be obtained whenever we
know the asymptotic of $\lambda_n$ and $\rho_n$ for large $n$. In
fact, if $\lambda_n\sim n^b$ and $\rho_n\sim n^a$, then the
general term of the associated zeta function behaves as
$n^{a-bs}$, and therefore the genus is
$q=\left[\frac{a+1}{b}\right]$ (the integer part).
\label{s3.r1}
\end{rem}

Some more remarks on these results are in order. First, note that the approach of considering some general class of abstract sequences and of studying the analytic properties of the associated spectral functions has been developed by various authors, and in particular instances of Proposition \ref{s3.p2} can be found in the literature. The original idea is probably due to Voros \cite{Vor}, while a good reference for a rigorous and very general setting is the work of Jorgenson and Lang \cite{JL}. However, for our purpose here, the simpler setting of \cite{Spr6} is more convenient.  Second, observe that Proposition \ref{product} was originally proved by Choi and Quine in \cite{CQ}, and also obtained in  \cite{Dow1}, equation (25). In particular, the reader can see the proof given in \cite{Spr6}, as the more rapid route to this result suggested in \cite{Dow1}.

\subsection{A class of  simple zeta functions}
\label{s3.2}

We consider the following class of simple zeta functions (compare
with \cite{Spr4})
\[
z(s;\alpha,\beta,x,p)=\sum_{n=1}^\infty
(n+x)^\alpha[(n+x)^\beta+p]^{-s},
\]
for $\Re(s)>\frac{1+\alpha}{\beta}$, where $\alpha$ and $\beta$
are real positive numbers,and $x$ and $p$ are real numbers
subject to the conditions that $n+x>0$ and $(n+x)^\beta+p>0$ for
all $n$.

Note that different equivalent techniques could be applied to
deal with this case; namely one could use the Plana theorem as in
\cite{Spr1}, a regularized product like in \cite{CQ}, a complex
integral representation as in \cite{Spr2}, or heat-kernel techniques
\cite{EORBZ} \cite{E}.

\begin{prop} The function $z(s;\alpha,\beta,x,p)$ has a regular analytic
continuation in the whole complex $s$-plane up to simple poles at
$s=\frac{1+\alpha}{\beta}-j$, $j=0,1,2,\dots$, when ever these
values are not $0,-1,-2,\dots$. The origin is a regular point and
if $\frac{1+\alpha}{\beta}$ is not a positive integer
\[
z(0;\alpha,\beta,x,p)= \zeta_H(-\alpha,x+1),
\]
and
\[
z'(0;\alpha,\beta,x,p)=\beta\zeta'_H(-\alpha,x+1)+\sum_{j=1}^{\left[\frac{\alpha+1}{\beta}\right]}
\frac{(-1)^j}{j}\zeta_H(\beta j-\alpha,x+1) p^j+
\]
\[
-\log\prod_{n=1}^\infty
\left(1+\frac{p}{(n+x)^\beta}\right)^{(n+x)^\alpha}\e^{(n+x)^\alpha\sum_{j=1}^{\left[\frac{\alpha+1}{\beta}\right]}
\frac{(-1)^j}{j}\frac{p^j}{(n+x)^{\beta j}}},
\]
while if $\frac{1+\alpha}{\beta}$ is a positive integer
\[
z(0;\alpha,\beta,x,p)= \zeta_H(-\alpha,x+1)
+\frac{(-1)^\frac{\alpha+1}{\beta}}{\alpha+1}p^\frac{\alpha+1}{\beta},
\]
and
\[
z'(0;\alpha,\beta,x,p)=\beta\zeta'_H(-\alpha,x+1)+\sum_{j=1}^{\frac{\alpha+1}{\beta}-1}
\frac{(-1)^j}{j}\zeta_H(\beta j-\alpha,x+1) p^j+
\]
\[
+\frac{(-1)^\frac{\alpha+1}{\beta}}{\frac{\alpha+1}{\beta}}
\left[-\Psi(x+1)+\left(\gamma+\Psi\left(\frac{1+\alpha}{\beta}\right)\right)\frac{1}{\beta}\right]p^\frac{\alpha+1}{\beta}+
\]
\[
-\log\prod_{n=1}^\infty
\left(1+\frac{p}{(n+x)^\beta}\right)^{(n+x)^\alpha}\e^{(n+x)^\alpha\sum_{j=1}^\frac{\alpha+1}{\beta}
\frac{(-1)^j}{j}\frac{p^j}{(n+x)^{\beta j}}}.
\]
\label{s3.p3}
\end{prop}
\begin{proof} The result follows applying Proposition \ref{s3.p2}.
First, note that the unshifted sequences is
$T_0=\{(n+x)^\beta\}$, with multiplicity $(n+x)^\alpha$. By the
Remark \ref{s3.r1}, the sequence has genus
$q=\left[\frac{\alpha+1}{\beta}\right]$. The associated zeta
function is
\[
z(s;\alpha,\beta,x,0)=\zeta_H(\beta s-\alpha,x+1),
\]
and this clearly shows that $T_0$ is a simply regular sequence of
spectral type, and so is $T_p$ by Lemma \ref{s3.l2}. The unique
pole is at $s=\frac{1+\alpha}{\beta}$ and
\[
{\rm Res}_1
\left(z(0;\alpha,\beta,x,0),s=\frac{1+\alpha}{\beta}\right)=\frac{1}{\beta},
\]
\[
{\rm Res}_0
\left(z(0;\alpha,\beta,x,0),s=\frac{1+\alpha}{\beta}\right)=
-\psi(x+1).
\]

The associated Fredholm determinant is
\[
F(z,T_0)=\prod_{n=1}^\infty
\left(1+\frac{z}{(n+x)^\beta}\right)^{(n+x)^\alpha}\e^{(n+x)^\alpha\sum_{j=1}^q
\frac{(-1)^j}{j}\frac{z^j}{(n+x)^{\beta j}}}.
\]

Next, using the expression given in the proof of Proposition
\ref{s3.p2} we have
\[
z(s;\alpha,\beta,x,p) = \sum_{j=0}^\infty\binom{-s}{j}
\zeta_H(\beta (s+j)-\alpha)p^j,
\]
thus we have poles when $\beta(s+j)-\alpha=1$, i.e.
$s=\frac{1+\alpha}{\beta}-j$, $j=0,1,2,\dots$, when ever these
values are not $0,-1,-2,\dots$, and the residua are easily
computed. To obtain the value at $s=0$, it is  useful to
distinguish two cases (see \cite{Spr4}). In fact, from the above
expression, when $s=0$ the unique term that is singular is the one
with $\beta j-\alpha=1$, i.e. $j=\frac{\alpha+1}{\beta}$, that is
necessary a positive integer since $\alpha\geq 0$. Now, if
$\frac{\alpha+1}{\beta}$ is not a positive integer, then we have
no integer poles,
$q={\left[\frac{\alpha+1}{\beta}\right]}\not=\frac{\alpha+1}{\beta}$,
and hence $z(0;\alpha,\beta,x,p)=z(0;\alpha,\beta,x,0)$, and since
\[
{\rm Res}_0
\left(z(s;\alpha,\beta,x,0),s=j\right)=z(j;\alpha,\beta,x,0)=\zeta_H(\beta
j-\alpha,x+1),
\]
\[
z'(0;\alpha,\beta,x,p)=z'(0;\alpha,\beta,x,0)
+\sum_{j=1}^{\left[\frac{\alpha+1}{\beta}\right]}
\frac{(-1)^j}{j}{\rm Res}_0
\left(z(s;\alpha,\beta,x,0),s=j\right)p^j 
-\log F(p,T_0).
\]

If $\frac{\alpha+1}{\beta}$ is a positive integer, we have a pole,
$q={\left[\frac{\alpha+1}{\beta}\right]}=\frac{\alpha+1}{\beta}$,
and we need to take in account also the residuum. As we have seen,
since the Hurwitz zeta function has only one pole at $s=1$ with
residuum 1, all the terms up to the ones with $j=0$ and the one
with $j=\frac{\alpha+1}{\beta}$ have vanishing residuum, and we
obtain
\[
z(0;\alpha,\beta,x,p)=z(0;\alpha,\beta,x,0)+\frac{(-1)^\frac{\alpha+1}{\beta}}{\frac{\alpha+1}{\beta}}{\rm
Res}_1
\left(z(s;\alpha,\beta,x,0),s=\frac{1+\alpha}{\beta}\right)p^\frac{\alpha+1}{\beta},
\]
and
\[
z'(0;\alpha,\beta,x,p)=z'(0;\alpha,\beta,x,0)+\sum_{j=1}^{\frac{\alpha+1}{\beta}-1}
\frac{(-1)^j}{j}{\rm Res}_0
\left(z(s;\alpha,\beta,x,0),s=j\right)p^j +
\]
\[
+\frac{(-1)^\frac{\alpha+1}{\beta}}{\frac{\alpha+1}{\beta}}\left[{\rm
Res}_0
\left(z(s;\alpha,\beta,x,0),s=\frac{1+\alpha}{\beta}\right)+
\right.
\]
\[
+\left.
\left(\gamma+\Psi\left(\frac{1+\alpha}{\beta}\right)\right){\rm
Res}_1
\left(z(s;\alpha,\beta,x,0),s=\frac{1+\alpha}{\beta}\right)\right]p^\frac{\alpha+1}{\beta}
-\log F(p,T_0),
\]
that gives the formula stated in the thesis.
\end{proof}

\subsection{A class of double zeta functions}
\label{s3.3}

Consider the following class of double zeta functions
\[
Z(s;\alpha,a,x,p)=\sum_{m,n=1}^\infty n^\alpha[(am+n+x)^2+p]^{-s},
\]
for $\Re(s)>1+\alpha$, and where $x$ and $p$ are real constants
subject to the  conditions that $am+n+x>0$ and $(am+n+x)^2+p>0$
for all $n$ and $m$, and $\alpha$ is a non negative integer (the
case where $\alpha$ is any real number can be treated by similar
methods, but is much more complicate, see \cite{Spr3}).

\begin{rem} In the more general case
\[
Z(s;\alpha,\beta,a,x,p)=\sum_{m,n=1}^\infty
n^\alpha[(am+n+x)^\beta+p]^{-s},
\]
for $\Re(s)>\frac{2(1+\alpha)}{\beta}$, and where $x$ and $p$ are
real constants subject to the  conditions that $am+n+x>0$ and
$(am+n+x)^\beta+p>0$ for all $n$ and $m$, we would have genus
$q=\left[\frac{2(1+\alpha)}{\beta}\right]$ by Remark \ref{s3.r1}
since the leading term behaves like $n^\alpha n^{-\beta s/2}$, but
we would not be able to prove that these are regular sequences of
spectral type as in the following proof of Lemma \ref{s3.l3}.
\label{s3.r2}
\end{rem}

The sequences appearing in these zeta functions are:
$S_0=\{\lambda_{m,n}=(am+n+x)^2\}_{m,n=1}^\infty$ and the
associated shifted sequence
$S_p=\{\lambda_{m,n}+p\}_{m,n=1}^\infty$, both with multiplicity
$n^\alpha$. These are sequences with finite exponent and genus
$q=\left[1+\alpha\right]$ by Remark \ref{s3.r1}. We first show
that $S_p$ is a simply regular sequence of spectral type.

\begin{lem}\label{s3.l3} The sequence $S_p=\{(am+n+x)^2+p\}_{m,n=1}^\infty$ is a simply regular
sequence of spectral type.
\end{lem}

\begin{proof} By Lemma \ref{s3.l2}, we need to show that there exists
an expansion of the desired type for the heat function
\[
f(t,S_0)=1+\sum_{m,n=1}^\infty
n^\alpha\e^{-(am+n+x)^2 t}.
\]

Consider the sequence $L=\{am+n+x\}_{m,n=1}^\infty$, with
multiplicity $n^\alpha$, of finite exponent and genus 2 (since
$m^a+n^b\leq (m n)^\frac{ab}{a+b}$). The associated heat function
is
\[
f(t,L)=1+\sum_{m,n=1}^\infty n^\alpha \e^{-(am+bn+c)t},
\]
and the associated Fredholm determinant is
\[
F(z,L)=\prod_{m,n=1}^\infty
\left(1+\frac{z}{am+bn+c}\right)^{n^\alpha}\e^{\sum_{j=1}^2
\frac{(-1)^j}{j}\frac{n^\alpha z^j}{(am+bn+c)^j}}.
\]

Since
\[
f(t,L)=1+\sum_{m,n=1}^\infty n^\alpha
\e^{-(am+bn+c)t}=1+\e^{-ct}\sum_{m=1}^\infty\e^{-amt}\sum_{n=1}^\infty
n^\alpha e^{-bnt},
\]
and we have an expansion of each factor in powers of $t$ (see
\cite{Spr4} Section 3.1 for the last sum), it is clear that we
have an expansion of the form
\[
f(t,L)=\sum_{j=0}^\infty e_j t^{\delta_j}.
\]

By Lemma 2.5 of \cite{Spr4}, $L$ is simply regular, and hence the
unique logarithmic terms in the expansion of $F(z,L)$ are of the
form $z^k \log z$, with integer $k\leq 2$. Now, consider the
product
\[
F(iz,L)F(-iz,L)=\prod_{m,n=1}^\infty
\left(1+\frac{iz}{am+bn+c}\right)^{n^\alpha}\left(1-\frac{iz}{am+bn+c}\right)^{n^\alpha}\times
\]
\[
\times\e^{\sum_{j=1}^2
\frac{(-1)^j}{j}\frac{{n^\alpha}(iz)^j}{(am+bn+c)^j}}\e^{\sum_{j=1}^2
\frac{(-1)^j}{j}\frac{{n^\alpha}(-iz)^j}{(am+bn+c)^j}}.
\]

Since $i^j+(-i)^j=0$ for odd $j$, and $-2$ when $j=2$, this gives
\[
F(iz,L)F(-iz,L)=\prod_{m,n=1}^\infty
\left(1+\frac{z^2}{(am+bn+c)^2}\right)^{n^\alpha}\e^{\frac{1}{2}
\frac{{n^\alpha}z^2}{(am+bn+c)^2}}=
F(z^2,S_0),
\]
and we obtain a decomposition of the Fredholm determinant
associated to the sequence $S_0$. This means that $\log F(z,S_0)$
has an expansion with unique logarithmic terms of the form
$z^k\log z$, with integer $k\leq 1$, and therefore $S_0$ is a
simply regular sequence of spectral type by Lemma 2.5 of
\cite{Spr6}.
\end{proof}

Lemma \ref{s3.l3} shows that the sequence appearing in the
definition of the function $Z(s;\alpha,a,x,p)=\zeta(s,S_p)$ are
such that we can apply Proposition \ref{s3.p2} in order to obtain
all the desired zeta invariants. For, we need explicit knowledge of
the zeta invariants of the sequence $S_0$. This is in the next
Lemma.

\begin{lem} The function $\chi(s;\alpha,a,x)$ defined for real
$a$ and $x$ such that $am+n+x>0$, for all $m,n\in \N_0$, and
$\alpha$ a non negative integer, by the sum
\[
\chi(s;\alpha,a,x)=\sum_{m,n=1}^\infty n^\alpha (am+n+x)^{-s},
\]
when $\Re(s)>2(\alpha+1)$, can be continued analytically to the
whole complex plane up to a finite set of simple poles at $s=1, 2,
\dots, \alpha+2$, by means of the following formula
\[
\chi(s;\alpha,a,x) =  \frac{1}{2}a^{-s}\zeta_H(s,(x+1)/a+1) +\frac{a^{1-s}}{s-1}\zeta_H(s-1,(x+1)/a+1)+
\]
\[
+\sum_{j=1}^\alpha \frac{\alpha (\alpha-1)\dots
(\alpha-j+1)}{(s-1)(s-2)\dots (s-j-1)} a^{j+1-s}
\zeta_H(s-j-1,(x+1)/a+1)  +
\]
\[
+ia^{-s}\int_0^\infty \frac{(1+i y)^\alpha \zeta_H(s,(x+1+i
y)/a+1)-(1-i y)^\alpha \zeta_H(s,(x+1-i y)/a+1)}{\e^{2\pi y}-1} d
y.
\]

In particular, this shows that the point $s=0$ is a regular point.
\label{s3.l4}
\end{lem}
\begin{proof} We apply the Plana theorem as in \cite{Spr1}.
Since the general term behaves as $n^\alpha n^{-s/2}$, we
assume $\Re(s)>2(\alpha+1)$.
\[
\chi(s;\alpha,a,x)=  \frac{1}{2}\sum_{m=1}^\infty (am+x+1)^{-s}
+\sum_{m=1}^\infty \int_1^\infty t^\alpha (am+t+x)^{-s} d t +
\]
\[
+i\sum_{m=1}^\infty\int_0^\infty \frac{(1+i y)^\alpha(am+x+1+i
y)^{-s}-(1-i y)^\alpha(am+x+1-i y)^{-s}}{\e^{2\pi y}-1} d y.
\]

Recall that $\alpha$ is a non negative integer, then we can
integrate recursively the middle term obtaining, for $\alpha>0$,
\[
\int_1^\infty t^\alpha (am+t+x)^{-s} d t=\sum_{j=0}^\alpha
\frac{\alpha (\alpha-1)\dots (\alpha-j+1)}{(s-1)(s-2)\dots
(s-j-1)} (am+x+1)^{j+1-s},
\]
this gives
\[
\chi(s;\alpha,a,x)=  \frac{1}{2}a^{-s}\sum_{m=1}^\infty
(m+(x+1)/a)^{-s} +\frac{a^{1-s}}{s-1}(m+(x+1)/a)^{1-s}+
\]
\[
+\sum_{j=1}^\alpha \frac{\alpha (\alpha-1)\dots
(\alpha-j+1)}{(s-1)(s-2)\dots (s-j-1)} a^{j+1-s}
\sum_{m=1}^\infty  (m+(x+1)/a)^{j+1-s}  +
\]
\[
+ia^{-s}\sum_{m=1}^\infty
\int_0^\infty \frac{(1+i y)^\alpha(m+(x+1+i y)/a)^{-s}-(1-i
y)^\alpha(m+(x+1-i y)/a)^{-s}}{\e^{2\pi y}-1} d y,
\]
and, due to uniform convergence of the integral, concludes the
proof.
\end{proof}

\begin{rem} We could deal with this kind of double zeta function
by applying the classical integral formula of Hermite as in the
case of the Riemann zeta function. This approach confirms the above results,
but it would not give a tractable expressions for the singular part.
\end{rem}

We can now obtain the zeta invariants of the zeta function
$Z(s;\alpha,a,x,p)$ for all the acceptable values of the parameters. This
allows us  to compute the regularized determinant of the deformed
sphere of any dimension, as pointed out at the beginning of this
section. Beside, we will give explicit formulas and results for
the low dimensional cases in the next subsections.

\subsection{Zeta determinant on the deformed 2 sphere}
\label{s3.4}

By Proposition \ref{s3.p0}, the zeta function associated to the
operator $\Delta_{S^2_{1/a}}$ is the function defined by the
series
\[
\zeta(s,\Delta_{S^2_{1/a}})=\sum_{n=1}^\infty
[n(n+1)]^{-s}+2\sum_{m=1,n=0}^\infty [(am+n)(am+n+1)]^{-s},
\]
when $\Re(s)>2$, and by analytic continuation elsewhere. The aim
of this section is to study this zeta function and in particular
to obtain a formula for the values of
$\zeta(0,\Delta_{S^2_{1/a}})$ and $\zeta'(0,\Delta_{S^2_{1/a}})$.
When $a=1$, this reduces to the zeta function on the 2-sphere: $
\zeta(s,\Delta_{S^2_{1/a=1}})= \sum_{n=1}^\infty
(2n+1)(n^2+n)^{-s}$ \cite{CQ} \cite{Spr1} \cite{Spr2}. The zeta
function $\zeta(s,S_{S_{1/a}^2})$ decompose as
\[
\zeta(s,\Delta_{S^2_{1/a}})=z(s;0,2,1/2,-1/4)+2Z(s;0,a,-1/2,-1/4),
\]
and we can easily check that the values of the parameters satisfy
the condition of definition of these functions. 
We provide two equivalent formulas for the zeta determinant on the deformed 2-sphere, Theorems \ref{s3.t1} and \ref{s3.t11}. The first is obtained applying Proposition \ref{s3.p2}, the second applying Proposition \ref{product}. Computations are given in the proofs of the following lemmas. The first lemma follows by a direct application  of Proposition \ref{s3.p3} and properties of special functions.

\begin{lem}
\[
z(0;0,2,1/2,-1/4)=-1,
\]
\[
z'(0;0,2,1/2,-1/4)=-\log 2 \pi.
\]
\label{s3.l5}
\end{lem}

\begin{lem}
\[
Z(0;0,a,-1/2,-1/4)=\frac{a}{12}+\frac{1}{12a},
\]
\[
Z'(0;0,a,-1/2,-1/4)=\frac{1}{6}\left(\frac{1}{2a}-a\right)\log a+
\]
\[
+\zeta_H'(0,1/(2a)+1)-2a\zeta_H(-1,1/(2a)+1)
-2a\zeta_H'(-1,1/(2a)+1) +
\]
\[
+2i\int_0^\infty \frac{\zeta_H'(0,(1/2+i y)/a+1)-\zeta_H'(0,(1/2-i
y)/a+1)}{\e^{2\pi y}-1} d y+
\]
\[
+\frac{1}{8a^2}\zeta_H(2,1/(2a)+1)-\frac{1}{4a}\left(\Psi(1/(2a)+1)+1+\log
a\right)+
\]
\[
 +\frac{i}{4a^2}\int_0^\infty \frac{\zeta_H(2,(1/2+i
y)/a+1)-\zeta_H(2,(1/2-i y)/a+1)}{\e^{2\pi y}-1}d y+
\]
\[
+\prod_{m,n=1}^\infty
\left(1-\frac{1}{4(am+n-1/2)^2}\right)\e^{\frac{1}{4(am+n-1/2)^2}}.
\]
\label{s3.l6}
\end{lem}
\begin{proof} The function $Z(s;0,a,-1/2,-1/4)$ is the
zeta function associated with the sequence
$S_{-1/4}=\{(am+n-1/2)^2-1/4\}$, all terms with multiplicity 1.
By Lemma \ref{s3.l3}, $S_{-1/4}$ is a simply regular sequence of
spectral type. In order to apply Proposition \ref{s3.p2}, we need
to study the unshifted sequence $S_0=\{(am+n-1/2)^2\}$. This
sequence has genus 1, the associate Fredholm determinant is
\[
F(z,S_0)=\prod_{m,n=1}^\infty
\left(1+\frac{z}{(am+n-1/2)^2}\right)\e^{-\frac{z}{(am+n-1/2)^2}},
\]
and the associated zeta function is
$\zeta(s,S_0)=\chi(2s;0,a,-1/2)$. By Proposition \ref{s3.p2} and
since the genus is 1, we have that
\[
Z(0;0,a,-1/2,-1/4)=\chi(0;0,a,-1/2)+\frac{1}{4}{\rm
Res}_1(\chi(2s;0,a,-1/2),s=1),
\]
and that
\[
\hspace{-0.7pt}Z'(0;0,a,-1/2,-1/4)\hspace{-0.3pt}=\hspace{-0.4pt}
\chi'(2s;0,a,-1/2)|_{s=0}+\frac{1}{4}\Res_0(\chi(2s;0,a,-1/2),s\hspace{-0.3pt}=\hspace{-0.3pt}1)-\log
F(-1/4,S_0),
\]
and hence we need to compute the values at $s=0$
of $\zeta(s,S_0)=\chi(2s;0,1,-1/2)$, and the residua at $s=1$.
For, we use the formula provided in  Lemma \ref{s3.l4}, namely
\[
\chi(2s;0,a,-1/2)= \frac{1}{2}a^{-2s}\zeta_H(2s,1/(2a)+1) +
\frac{1}{2s-1} a^{1-2s} \zeta_H(2s-1,1/(2a)+1)  +
\]
\beq +ia^{-2s}\int_0^\infty \frac{\zeta_H(2s,(1/2+i
y)/a+1)-\zeta_H(2s,(1/2-i y)/a+1)}{\e^{2\pi y}-1} d y. \label{e1}
\eeq

We obtain
\[
\chi(0;0,a,-1/2)= \frac{1}{2}\zeta_H(0,1/(2a)+1) -a
\zeta_H(-1,1/(2a)+1)  +
\]
\[
+i\int_0^\infty \frac{\zeta_H(0,(1/2+i y)/a+1)-\zeta_H(0,(1/2-i
y)/a+1)}{\e^{2\pi y}-1} d y= \frac{a}{12}-\frac{1}{24a},
\]
where we have used \cite{GZ} 9.531 and 9.611.1. Next, we use
equation (\ref{e1}) to compute the residua at the pole $s=1$. The
unique singular term is the middle one, so we expand the different
factors in it near $s=1$, using \cite{GZ} 9.533.2,
\[
a \frac{a^{-2s}}{2s-1}
\zeta_H(2s-1,1+1/(2a))=\frac{1}{2a}\frac{1}{s-1}-\frac{1}{a}\left(\Psi(1+1/(2a))+1+\log
a\right)+O(s-1).
\]

This gives
\[
\Res_1(\chi(2s;0,a,-1/2),s=1)=\frac{1}{2a},
\]
and

\[
\Res_0(\chi(2s;0,a,-1/2),s=1)=
\frac{1}{2a^2}\zeta_H(2,1/(2a)+1)-\frac{1}{a}\left(1+\log
a\right)+
\]
\[
-\frac{1}{a}\Psi(1/(2a)+1)+ \frac{i}{a^2}\int_0^\infty
\frac{\zeta_H(2,(1/2+i y)/a+1)-\zeta_H(2,(1/2-i y)/a+1)}{\e^{2\pi
y}-1}d y.
\]

Last, we compute the derivative:
\[
\chi'(0;0,a,-1/2)= -2\chi(0;0,a,-1/2)\log a+
\]
\[
+\zeta_H'(0,1/(2a)+1)-2a\zeta_H(-1,1/(2a)+1)
-2a\zeta_H'(-1,1/(2a)+1) +
\]
\[
+2i\int_0^\infty \frac{\zeta_H'(0,(1/2+i y)/a+1)-\zeta_H'(0,(1/2-i
y)/a+1)}{\e^{2\pi y}-1} d y=
\]
\[
=\frac{1}{6}\left(\frac{1}{2a}-a\right)\log
a+\Gamma(1/(2a)+1)-\frac{1}{2}\log
2\pi+\frac{a}{6}+\frac{1}{4a}+\frac{1}{2}-2a\zeta_H'(-1,1/(2a)+1)+
\]
\[
+ 2i\int_0^\infty
\log\frac{\Gamma((1/2+iy)/a+1)}{\Gamma((1/2-iy)/a+1)}\frac{dy}{\e^{2\pi
y}-1} d y.
\]

Collecting, we obtain the thesis.
\end{proof}

\begin{lem}
\[
Z'(0;0,a,-1/2,-1/4)=-\left(\frac{a}{6}+\frac{1}{6a}\right)\log a
-\frac{1}{2}\log 2\pi
+\frac{1}{2}\log \Gamma(1+\frac{1}{a})+
\]
\[
+\frac{a}{6}+\frac{1}{2}+\frac{3}{4a}
-a\zeta_R'(-1)-a\zeta_H'(-1,1+\frac{1}{a})+
\]
\[
+i\int_0^\infty \log \frac{\Gamma(1+i\frac{y}{a})\Gamma(1+\frac{1}{a}+i\frac{y}{a})}{\Gamma(1-i\frac{y}{a})\Gamma(1+\frac{1}{a}-i\frac{y}{a})}\frac{d y}{\e^{2\pi y}-1}.
\]

\label{s3.l66}
\end{lem}
\begin{proof} 
In the language of Proposition \ref{product}, we have
\[
L_0=\{(am+n+x)^2\}, \hspace{30pt} L_{b^2}=\{(am+n+x)^2+b^2\},
\]
\[
S_0=\{am+n+x\}, \hspace{30pt} S_{ib}=\{am+n+x+ib\},
\]
where the genus of $S_0$ is $p=2$. Therefore, by proposition \ref{product},
\[
Z'(0;0,a,x,b^2)=\zeta'(0,L_{b^2})=\zeta'(0,S_{ib})+\zeta'(0,S_{-ib}))-\Res_1(\zeta(s,S_0),s=2) b^2. 
\]
Also, we have that
\[
\zeta(s,S_{ib})=\sum_{m,n=1}^\infty (am+n+x+ib)^{-s}=\chi(s;0,a,x+ib),
\]
and therefore, we need information on $\chi$. Use Lemma \ref{s3.l4}. We have, with $z=x\pm ib$,
\[
\chi(s;0,a,z) =  \frac{1}{2}a^{-s}\zeta_H(s,\frac{z+1}{a}+1) +
\frac{a^{1-s}}{s-1}\zeta(s-1,\frac{z+1}{a}+1) +
\]
\[
+ia^{-s}\int_0^\infty \frac{\zeta_H(s,\frac{z+1 +i
y}{a}+1)- \zeta_H(s,\frac{z+1 -i y}{a}+1)}{\e^{2\pi y}-1} d
y.
\]

This gives 
\[
\Res_1(\chi(s;0,a,z),s=2)=\frac{1}{a},
\]
\[
\chi(0;0,a,z)=\frac{1}{4}+\frac{a}{12}+\frac{1}{12a}+\frac{z^2}{2a}+\frac{z}{2a}+\frac{z}{2},
\]
\[
\chi'(0;0,a,z)=-\chi(0;0,a,z)\log a
+\frac{1}{2}\zeta_H'(0,\frac{z+1}{a}+1)
-a\zeta_H(-1,\frac{z+1}{a}+1)+
\]
\[
-a\zeta_H'(-1,\frac{z+1}{a}+1)
+i\int_0^\infty \log \frac{\Gamma(1+\frac{z+iy+1}{a})}{\Gamma(1+\frac{z-iy+1}{a})}\frac{d y}{\e^{2\pi y}-1}.
\]

\end{proof}

Using the decomposition at the beginning of this subsection and
the results in Lemmas \ref{s3.l5}, \ref{s3.l6} and \ref{s3.l66} respectively, we can prove the
following theorems.

\begin{teo}\label{s3.t1}
\[
\zeta(0,\Delta_{S^2_{1/a}})=-1+\frac{a}{6}+\frac{1}{6a},
\]
\[
\zeta'(0,\Delta_{S^2_{1/a}})=-2\log
2\pi+1+\frac{a}{3}-\frac{1}{3}\left(a+\frac{1}{a}\right)\log
a+2\log\Gamma(1/(2a)+1)+
\]
\[
+\frac{1}{4
a^2}\zeta_H(2,1/(2a)+1)-\frac{1}{2a}\Psi(1/(2a)+1)-4a\zeta_H'(-1,1/(2a)+1)+
\]
\[
+4i\int_0^\infty \log\frac{\Gamma((1/2+i y)/a+1)}{\Gamma((1/2-i
y)/a+1)}\frac{d y}{\e^{2\pi y}-1}+
\]
\[
+\frac{i}{2a^2}\int_0^\infty \frac{\zeta_H(2,(1/2+i
y)/a+1)-\zeta_H(2,(1/2-i y)/a+1)}{\e^{2\pi y}-1}d y+
\]
\[
-2\log \prod_{m,n=1}^\infty
\left(1-\frac{1}{4(am+n-1/2)^2}\right)\e^{\frac{1}{4(am+n-1/2)^2}}.
\]
\end{teo}

\begin{teo}\label{s3.t11}

\[
\zeta'(0,\Delta_{S^2_{1/a}})=-\left(\frac{a}{3}+\frac{1}{3a}\right)\log a
-2\log 2\pi
+\frac{a}{3}+1+\frac{3}{2a}+\log \Gamma(1+\frac{1}{a})+
\]
\[
-2a\zeta_R'(-1)-2a\zeta_H'(-1,1+\frac{1}{a})+
2i\int_0^\infty \log \frac{\Gamma(1+i\frac{y}{a})\Gamma(1+\frac{1}{a}+i\frac{y}{a})}{\Gamma(1-i\frac{y}{a})\Gamma(1+\frac{1}{a}-i\frac{y}{a})}\frac{d y}{\e^{2\pi y}-1}.
\]

\end{teo}

Observe that, although the formula given in Theorem \ref{s3.t11} looks nicer, it is in fact less useful than the one given in Theorem \ref{s3.t1}, since convergence of the integral is much lower that convergence of the infinite product. 
Note also that the analytic formulas obtained in the previous theorems, provides a rigorous answer to the problem studied in \cite{Dow3}, where an attempt to obtain such formulas was performed. In particular, we can compare the graphs given in \cite{Dow3} Section XI (where observe the opposite sign), with the following one, where $\zeta'(0,\Delta_{S^2_{1/a}})$ is plotted using the formula given in Theorem \ref{s3.t1}, and the relation with the lune angle $\omega$ is $a=\frac{\pi}{\omega}$.

\begin{center}
\includegraphics{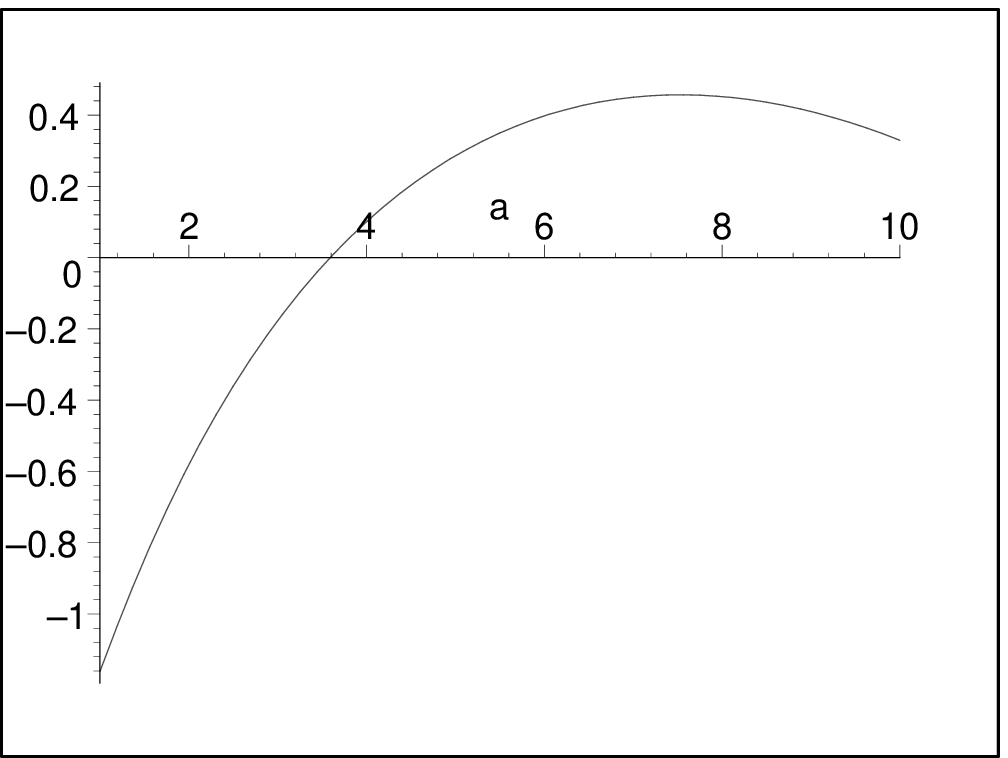}
\end{center}

\subsection{The zeta determinant on the deformed 3 sphere}
\label{s4}

On the deformed 3-sphere we have $N=2$ and
\[
\zeta(s,\Delta_{ S^{3}_{1/a}})=\sum_{n=1}^\infty
(n+1)[n(n+2)]^{-s}+2\sum_{m=1}^\infty \sum_{n=0}^\infty
(n+1)[(am+n)(am+n+2)]^{-s}.
\]

We can check that this reduces to the usual zeta function on the
3-sphere $\zeta(s,\Delta_{ S^{3}_1})= \sum_{n=1}^\infty
(n+1)^2[n(n+2)]^{-s}$ \cite{Spr1} \cite{CQ}, and we can decompose
it as follows
\[
\zeta(s,\Delta_{ S^{3}_{1/a}})=z(s;1,2,1,-1)+2Z(s;1,a,0,-1).
\]

As in the previous subsection, we apply Propositions \ref{s3.p2} and \ref{product}
and properties of special functions to prove the following lemmas. Observe that, in this case, an application of Proposition \ref{product} gives a simpler formula for $z'(0;1,2,1,-1)$, we thanks the referee for pointing out this fact.

\begin{lem}
\[
z(0;1,2,1,-1)=-1,
\]
\[
z'(0;1,2,1,-1)=\gamma-1+2\zeta_H'(-1,2)-\log \prod_{n=2}^\infty
\left(1-\frac{1}{n^2}\right)^n \e^\frac{1}{n}=2\zeta'_H(-1,2)+\log 2-1.
\]
\label{s3.l7}
\end{lem}

\begin{rem} The above result allows to obtain the following interesting formulas for the Barnes G-function $G(z)$ and the double sine function $S(z)$ (see \cite{Bar1}, \cite{Shu} or \cite{Spr6} for the definition of the G-function, and \cite{Kur} or \cite{Spr6} for the multiple sine function):
\[
\lim_{z\to 1} \frac{G(1-z)}{1-z}=\frac{4}{\e},
\]
\[
\lim_{z\to 1} \frac{S(\pi(1-z))}{1-z}=\frac{\pi}{\e}.
\]
\end{rem}

The proofs of the next lemmas go as the one of Lemmas \ref{s3.l6} and \ref{s3.l66}.
Beside the increasing difficulty of the calculation and the fact
that now the multiplicity is not trivial ($\alpha=1$), the main difference is
that a new singular term appears in the unshifted zeta function,
namely applying Lemma \ref{s3.l4}, we obtain the expression
\[
\chi(2s;1,a,0)= \frac{a^{-2s}}{2}\zeta_H(2s,1/a+1)
+\frac{a^{1-2s}}{2s-1}  \zeta_H(2s-1,1/a+1)  +
\]
\[
+ \frac{a^{2-2s}}{(2s-1)(2s-2)}  \zeta_H(2s-2,1/a+1)  +
\]
\[
+ia^{-2s}\int_0^\infty \frac{(1+i y)\zeta_H(2s,(1+i y)/a+1)-(1-i
y)\zeta_H(2s,(1-i y)/a+1)}{\e^{2\pi y}-1} d y.
\]
instead of formula (\ref{e1}).

\begin{lem}
\[
Z(0;1,a,0,-1)=-\frac{5}{24},
\]
\[
Z'(0;1,a,0,-1)=
\frac{3}{4}-\frac{a}{12}+\frac{1}{2a}+\frac{5}{12}\log
a-\frac{11}{12}\log 2\pi+ 2\log\Gamma(1/a+1)+
\]
\[
+\frac{1}{2a^2}\zeta_H(2,1/a+1)-\frac{1}{a}\Psi(1+1/a) -2a
\zeta_H'(-1,1/a+1) + a^2\zeta_H'(-2,1/a+1)+
\]
\[
+2i\int_0^\infty \log\frac{\Gamma((1+i y)/a+1)}{\Gamma((1-i
y)/a+1)}\frac{d y}{\e^{2\pi y}-1}- 2\int_0^\infty
y\log|\Gamma((1+i y)/a+1)|^2 \frac{d y}{\e^{2\pi y}-1}+
\]
\[
+\frac{i}{a^2}\int_0^\infty \frac{\zeta_H(2,(1+i
y)/a+1)-\zeta_H(2,(1-i y)/a+1)}{\e^{2\pi y}-1}d y+
\]
\[
-\frac{1}{a^2}\int_0^\infty y\frac{\zeta_H(2,(1+i
y)/a+1)+\zeta_H(2,(1-i y)/a+1)}{\e^{2\pi y}-1}d y+
\]
\[
-\log\prod_{m,n=1}^\infty
\left(1-\frac{1}{(am+n)^2}\right)^n\e^{\frac{n}{(am+n)^2}}.
\]
\label{s3.l8}
\end{lem}

\begin{lem}\label{s3.l88}
\[
Z'(0;1,a,0,-1)=
\frac{5}{12}\log a -1-\frac{a}{12}-\frac{5}{12}\log 2\pi+\frac{1}{2}\log \Gamma( \frac{2}{a}+1)+
\]
\[
-a\left(\zeta_R'(-1)+\zeta_H'(-1,\frac{2}{a}+1)\right)
+\frac{a^{2}}{2}\left(\zeta_R'(-2)+\zeta_H'(-2,\frac{2}{a}+1)\right)+
\]
\[
+i\int_0^\infty \log\frac{\Gamma(1+i\frac{y}{a})\Gamma(1+\frac{2+iy}{a})}{\Gamma(1-i\frac{y}{a})\Gamma(1+\frac{2-iy}{a})}\frac{dy}{\e^{2\pi y}-1}+
\]
\[
-\int_0^\infty y\log \frac{\pi y}{a{\rm sh}\frac{\pi y}{a}}\Gamma(1+\frac{2+iy}{a})\Gamma(1+\frac{2-iy}{a})\frac{dy}{\e^{2\pi y}-1}.
\]

\end{lem}

Using the decomposition at the beginning of this subsection and
the results in Lemmas \ref{s3.l7}, \ref{s3.l8} and \ref{s3.l88} we can prove the
following theorems.

\begin{teo}\label{s3.t2}
\[
\zeta(0,\Delta_{S^3_{1/a}})=-1,
\]
\[
\zeta'(0,\Delta_{S^3_{1/a}})=\gamma-1+2\zeta_H'(-1,2)-\log
\prod_{n=2}^\infty \left(1-\frac{1}{n^2}\right)^n \e^\frac{1}{n}+
\]
\[
+\frac{3}{2}-\frac{a}{6}+\frac{1}{a}+\frac{5}{6}\log
a-\frac{11}{6}\log 2\pi+ 4\log\Gamma(1/a+1)+
\]
\[
+\frac{1}{a^2}\zeta_H(2,1/a+1)-\frac{2}{a}\Psi(1+1/a) -4a
\zeta_H'(-1,1/a+1) + 2a^2\zeta_H'(-2,1/a+1)+
\]
\[
+4i\int_0^\infty \frac{(1+i y)\log\Gamma((1+i y)/a+1)-(1-i
y)\log\Gamma((1-i y)/a+1)}{\e^{2\pi y}-1}d y+
\]
\[
+\frac{2i}{a^2}\int_0^\infty \frac{(1+i y)\zeta_H(2,(1+i
y)/a+1)-(1-i y)\zeta_H(2,(1-i y)/a+1)}{\e^{2\pi y}-1}d y+
\]
\[
-2\log\prod_{m,n=1}^\infty
\left(1-\frac{1}{(am+n)^2}\right)^n\e^{\frac{n}{(am+n)^2}}.
\]
\end{teo}

\begin{teo} \label{s3.t22}
\[
\zeta'(0,\Delta_{S^3_{1/a}})=\log 2+2\zeta'_R(-1)-1+
\frac{5}{6}\log a -2-\frac{a}{6}-\frac{5}{6}\log 2\pi+\log \Gamma( \frac{2}{a}+1)+
\]
\[
-2a\left(\zeta_R'(-1)+\zeta_H'(-1,\frac{2}{a}+1)\right)
+a^{2}\left(\zeta_R'(-2)+\zeta_H'(-2,\frac{2}{a}+1)\right)+
\]
\[
+2i\int_0^\infty \log\frac{\Gamma(1+i\frac{y}{a})\Gamma(1+\frac{2+iy}{a})}{\Gamma(1-i\frac{y}{a})\Gamma(1+\frac{2-iy}{a})}\frac{dy}{\e^{2\pi y}-1}
-2\int_0^\infty y\log \frac{\pi y\Gamma(1+\frac{2+iy}{a})\Gamma(1+\frac{2-iy}{a})}{a{\rm sh}\frac{\pi y}{a}}\frac{dy}{\e^{2\pi y}-1}.
\]
\end{teo}

\subsection{Expansions}

In this subsection we give explicit formulas and numerical values
of the first coefficients appearing in the expansions of the
determinants of the Laplace operator on the 2 and 3 dimensional
deformed sphere $S_k^N$ for small deformations of the parameter
$k=1-\delta$, with small positive $\delta$. We first state a lemma
that allows to deal with the expansion of the values of the zeta
function, and thus justify the formal series expansion of all the
functions appearing in Theorems \ref{s3.t1} and \ref{s3.t2} up to
the infinite products, but the last can be treated directly. The
proof of Lemma \ref{a1.l1} follows by the same argument as the one
used in the proof of Proposition \ref{s3.p2}.

\begin{lem} Let $x,q$ and $\delta$ be real with $0\leq\delta\leq
1$, then for all $\Re(s)>-2$ we have the expansion
\[
\zeta_H(s,1+x+q\delta))
=\zeta_H(s,1+x)-s\zeta_H(s+1,1+x)q\delta+\frac{s(s+1)}{2}\zeta_H(s+2,1+x)q^2\delta^2
+O(\delta^3),
\]
and
\[
\zeta'_H(s,1+x+q\delta))
=\zeta'_H(s,1+x)-\left(\zeta_H(s+1,1+x)+s\zeta'_H(s+1,1+x)\right)q\delta+
\]
\[
+\left(\left(s+\frac{1}{2}\right)\zeta_H(s+2,1+x)+\frac{s(s+1)}{2}\zeta'_H(s+2,1+x)\right)q^2\delta^2
+O(\delta^3),
\]
where note that the coefficients of the second and third term in
the second formula are defined as limits. \label{a1.l1}
\end{lem}

\begin{prop} For $a=1+\delta+O(\delta^2)$,
\[
\zeta'(0,\Delta_{S^2_{1-\delta}})=\zeta'(0,\Delta_{S^2_1})+Z_2
\delta+O(\delta^2),
\]
where
\[
\zeta'(0,\Delta_{S^2_1})=4\zeta'_H(-1)-\frac{1}{2}=
\]
\[
=-\log2
\pi-\frac{2}{3}+\frac{\pi}{8}+\frac{\gamma}{2}-4\zeta'_H(-1,1/2)+\frac{i}{2}\int_0^\infty
\frac{\Psi'(3/2+i y)-\Psi'(3/2-i y)}{\e^{2\pi y}-1}d y+
\]
\[
+4i\int_0^\infty \log\frac{\Gamma(3/2+i y)}{\Gamma(3/2-i
y)}\frac{d y}{\e^{2\pi y}-1}-2\log \prod_{m,n=1}^\infty
\left(1-\frac{1}{4(m+n+1/2)^2}\right)\e^{\frac{1}{4(m+n+1/2)^2}}
\]
\[
=-1.161684575,
\]
\[
Z_2=-\frac{1}{3}+\frac{\gamma}{2}-\frac{\pi^2}{8}+\frac{7}{4}\zeta_R(3)-4\zeta'_H(-1,1/2)+
2\pi\int_0^\infty \frac{\tanh \pi y}{\e^{2\pi y}-1}d y+
\]
\[
+4\int_0^\infty y\frac{\Psi(1/2+i y)+\Psi(1/2-i y)}{\e^{2\pi
y}-1}d y +\frac{1}{2}\int_0^\infty y\frac{\Psi''(3/2+i
y)+\Psi'(3/2-i y)}{\e^{2\pi y}-1}d y+
\]
\[
-\frac{i}{4}\int_0^\infty \frac{\Psi''(3/2+i y)-\Psi'(3/2-i
y)}{\e^{2\pi y}-1}d y -4\sum_{j=2}^\infty
\frac{1}{4^j}\sum_{m,n=1}^\infty \frac{m}{(m+n+1/2)^{2j+1}}=
\]
\[
=0.7116523492.
\]
\end{prop}

\begin{corol}
\[
{\rm det}\Delta_{S^2_{1-\delta}}={\rm det}\Delta_{S^2_{1}}-Z_2{\rm
det}\Delta_{S^2_{1}}\delta+O(\delta^2)=3.195311305-2.273950797
\delta+O(\delta^2).
\]
\end{corol}

\begin{prop} For $a=1+\delta+O(\delta^2)$,
\[
\zeta'(0,\Delta_{S^3_{1-\delta}})=\zeta'(0,\Delta_{S^3_1})+Z_3
\delta+O(\delta^2),
\]
where
\[
\zeta'(0,\Delta_{S^3_1})=2\zeta'_R(-2)+2\zeta_R'(0)+\log 2=
\]
\[
=3\gamma-\frac{5}{3}-2\zeta_R'(-1)-\frac{11}{6}\log(2\pi)+\frac{\pi^2}{6}+2\zeta_R'(-2)-\log
\prod_{n=2}^\infty \left(1-\frac{1}{n^2}\right)^n \e^\frac{1}{n}+
\]
\[
+4i\int_0^\infty \log\frac{\Gamma(2+i y)}{\Gamma(2-i y)}\frac{d
y}{\e^{2\pi y}-1} - 4\int_0^\infty y\log|\Gamma(2+i y)|^2 \frac{d
y}{\e^{2\pi y}-1}+
\]
\[
+ 2i\int_0^\infty \frac{\zeta_H(2,2+i y)-\zeta_H(2,2-i
y)}{\e^{2\pi y}-1}d y -2\int_0^\infty y\frac{\zeta_H(2,2+i
y)+\zeta_H(2,2-i y)}{\e^{2\pi y}-1}d y+
\]
\[
=-2\log\prod_{m,n=1}^\infty
\left(1-\frac{1}{(m+n)^2}\right)^n\e^{\frac{n}{(m+n)^2}}
=-1.205626800,
\]
\[
Z_3=-\frac{1}{2}+2\gamma+2\zeta_R(3)-8\zeta'_R(-1)-2\log(2
\pi)+4\zeta_R'(-2)+
\]
\[
-4i\int_0^\infty \frac{(1+i y)^2\Psi(2+i y)-(1-i y)^2\Psi(2-i
y)}{\e^{2\pi y}-1}d y+
\]
\[
-4i\int_0^\infty \frac{(1+i y)\Psi'(2+i y)-(1-i y)\Psi'(2-i
y)}{\e^{2\pi y}-1}d y+
\]
\[
-2i\int_0^\infty \frac{(1+i y)^2\Psi''(2+i y)-(2-i y)^2\Psi''(2-i
y)}{\e^{2\pi y}-1}d y +\frac{3}{2}-\frac{\pi^2}{9}
=0.6666666661=\frac{2}{3}.
\]
\end{prop}

\begin{corol}
\[
{\rm det}\Delta_{S^3_{1-\delta}}={\rm det}\Delta_{S^3_{1}}-Z_3{\rm
det}\Delta_{S^3_{1}}\delta+O(\delta^2)=
3.338845845-2.225897228\delta+O(\delta^2).
\]
\end{corol}

{\bf Acknowledgments}

We would like to thank an anonymous referee for useful remarks and suggestions.
 One of the author, M. S., thanks the Departments of Mathematics
and Physics of the University of Trento, and the INFN for nice
hospitality. S. Z. thanks V. Moretti for discussions.

\end{document}